\documentclass[doublecol]{epl2} 
\usepackage{graphicx}
\usepackage{amsmath}
\usepackage{comment}
\usepackage{subfigure}

\title{Quantum Capacitance Modifies Interionic Interactions in Semiconducting Nanopores}
\author{Alpha A. Lee\inst{1,2} \and Dominic Vella \inst{1} \and Alain Goriely\inst{1}}
\shortauthor{A. A. Lee \etal}

\institute{                    
  \inst{1} Mathematical Institute, University of Oxford, Woodstock Rd, Oxford, OX2 6GG, UK\\
  \inst{2} John A. Paulson School of Engineering and Applied Sciences, Harvard University, Cambridge, MA 02138
}

\abstract{
Nanopores made with low dimensional semiconducting materials, such as carbon nanotubes and graphene slit pores, are used in supercapacitors. In theories and simulations of their operation, it is often assumed that such pores screen ion-ion interactions like metallic pores, i.e. that screening leads to an exponential decay of the interaction potential with ion separation. By introducing a quantum capacitance that accounts for the density of states in the material, we show that ion-ion interactions in carbon nanotubes and graphene slit pores actually decay algebraically with ion separation. This result suggests a new avenue of capacitance optimization based on tuning the electronic structure of a pore: a marked enhancement in capacitance might be achieved by developing nanopores made with metallic materials or bulk semimetallic materials. }

\begin{document}

\maketitle

\section{Introduction}

Confinement of ions in nanostructures underpins the physics of many new electrochemical devices, ranging from supercapacitors \cite{brandt2013ionic} to field effect transistors \cite{fujimoto2013electric}. These nanostructures are usually porous semiconducting materials such as carbon nanotubes or graphene slit pores. Early experiments have shown that electrodes made with porous carbide-derived-carbon deliver large volumetric capacitance as the pore-size (which can be precisely controlled) approaches the ion size \cite{chmiola2006anomalous, largeot2008relation,lin2009solvent}. This increase in capacitance cannot be rationalised by surface area enhancement alone, leading to the general hypothesis that the electronic structure of the electrodes significantly modifies the Coulomb interaction between ions \cite{kondrat2011superionic1,skinner2011theory}. 

Most theoretical studies of supercapacitors to date have modelled the electrode material as ideally metallic \cite{kondrat2011superionic1,kondrat2011superionic,skinner2011theory, kondrat2014single}, and studied the effect of ion size and pore size on the capacitance. The key assumption in ideal metal theory is that the electric field cannot penetrate into the bulk metal: the high concentration of electrons in metals screens the electric field completely, and only a surface charge density is induced. In particular, for a charge positioned in front of an infinite metal slab, the induced surface charge density is equivalent to an equal and opposite ``image'' charge located inside the metal at the same distance away from the metal surface \cite{jackson1962classical}. In a metallic pore, this ion-image interaction modifies the long-ranged $\propto 1/r$ Coulomb interaction energy to an exponentially decaying interaction energy. The characteristic screening length is the pore width or diameter. 

For semi-metallic materials, the effect of finite electron concentration has been studied in \cite{skinner2011theory,skinner2011model,rochester2013interionic} using the Thomas-Fermi model, which accounts for the effect of electric field penetration into the material. It was shown that for a slit/cylindrical pore, the pore width/diameter is effectively renormalised by the Thomas-Fermi screening length but that the decay remains exponential. However, the concept of electric field penetration is only physical for bulk materials and not applicable for 2D materials such as single-layer graphene slit pores and carbon nanotubes, where the material is only one atom thick. Nonetheless, a phenomenological approach based on introducing an effective pore radius or width has been successfully used in \cite{mohammadzadeh2014nanotubes} to fit detailed simulation data for the interaction potential between ions in single-layer gold and carbon nanotubes. 

In this paper, we model the interionic interactions in a graphene slit pore or carbon nanotube by introducing the quantum capacitance, a key quantity capturing the quantum density-of-states of a material. Analytical expressions can be obtained in the linear-response regime. Surprisingly, a finite quantum capacitance modifies the electrostatic interactions between ions: rather than being exponentially decaying, they are algebraic $\propto 1/r^3$ for a slit pore and $\propto 1/(r \log^2 r)$ for a cylindrical pore. Our results demonstrate a fundamental quantum limitation of using carbon nanotubes or graphene slit pores as materials for supercapacitors, and suggest that metallic or bulk semimetallic materials may be more suitable for supercapacitor applications. 

\section{Electrostatic Interactions and Quantum Capacitance}
\begin{figure}
\centering
\subfigure[]{\includegraphics[scale=0.35]{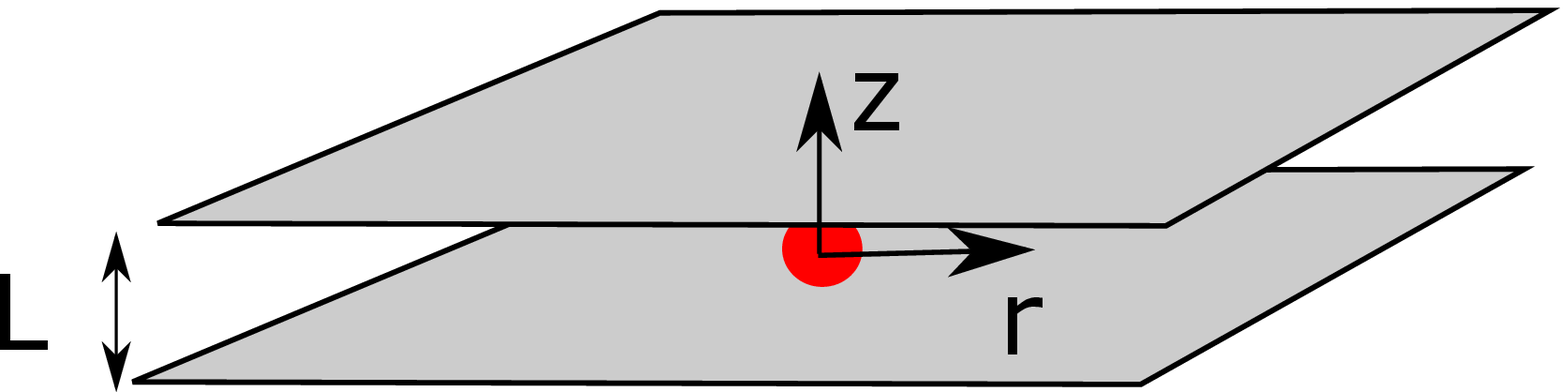}}
\subfigure[]{\includegraphics[scale=0.35]{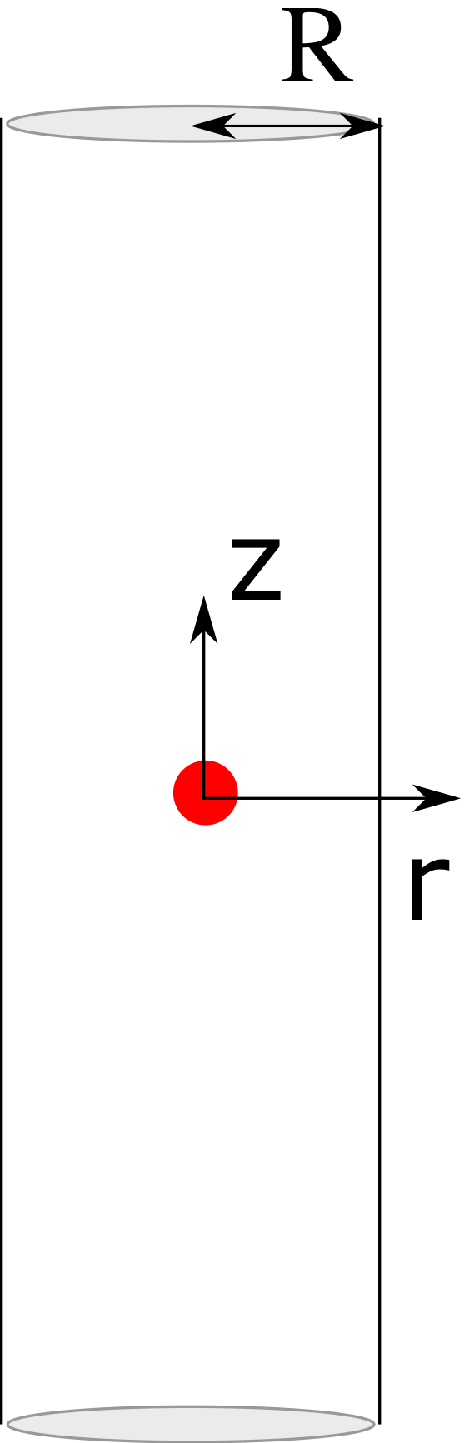}}
\caption{Sketch of the systems under consideration, a positive ion of charge $q$ positioned on (a) central symmetry plane of a graphene slit pore of width $L$, and (b) the central symmetry axis of a nanotube with radius $R$.}
\label{schematic_nanotube}
\end{figure}

We consider two idealised geometries of supercapacitors: a carbon nanotube of radius $R$ and a 2-D graphene slit pore of width $L$. Both geometries are idealisations of ``real'' porous materials such as carbide-derived-carbon; these are typically disordered with domains resembling a slit/cylindrical pore. However, these idealizations allows us to make analytical progress and yield more insight. In each geometry, a charge $q$ is positioned on the central symmetry axis/plane (see Figure \ref{schematic_nanotube}). 
The interaction energy between a test charge with the same charge $q$ located at $(r,z)$ and a fixed point charge is given by 
\begin{equation}
U(r,z) = q \phi(r,z), 
\end{equation}
where $\phi$ is the electric potential induced by the point charge. 

To calculate the electric potential $\phi$, we need to couple the classical electrostatic boundary value problem \cite{jackson1962classical} with quantum capacitance. The quantum capacitance arises as one of the electrostatic boundary conditions: the electric field inside and outside a tube/slit pore is related to the induced surface charge $\sigma$ via the Gauss law
\begin{equation}
\epsilon [\mathbf{n} \cdot \nabla \phi] = - 4 \pi \sigma,
\label{G_law}
\end{equation} 
where $[.]$ denotes the jump across the tube/slit surface, $\mathbf{n}$ is the unit vector normal to the surface (defined to point in the direction outside of the pore), and $\epsilon$ is the relative dielectric constant of the medium; for simplicity we assume $\epsilon$ to be the same inside and outside the pore. 

The induced surface charge is a function of the electric potential on the pore surface $\phi_s(\mathbf{r})$, i.e. $\sigma = \sigma(\phi_s)$. In particular, within a local density approximation \cite{ghaznavi2010nonlinear} 
\begin{equation} 
\sigma(\phi_s) = -e \left[ n(\mu + e \phi_s) - n (\mu) \right], 
\label{TF_nonlin}
\end{equation}
where $n$ is the density of charge carriers and $\mu$ is the chemical potential. The chemical potential is changed through doping, with $\mu>0 \; (\mu<0)$ for electron (hole) doping. Analytical approximations for $n(x)$ have been proposed for doped graphene \cite{ghaznavi2010nonlinear}. For simplicity, however, we shall assume that the induced surface potential $ e \phi_s \ll \mu$, and linearise Equation (\ref{TF_nonlin}) to obtain
\begin{equation} 
%\sigma(\phi_s) \approx -e^2 n'(\mu) \phi_s \equiv - \frac{C_q}{4 \pi} \phi_s. 
\sigma(\phi_s) \approx -e^2 n'(\mu) \phi_s \equiv - \frac{\epsilon}{4\pi} C_q\phi_s. 
\label{q_cap_lin}
\end{equation}
%Here, we have identified $C_q =  4 \pi e^2 n'(\mu)/\epsilon$ as the linear quantum capacitance. Therefore, the Gauss Law (\ref{G_law}) becomes a Robin boundary condition  
Here, we have identified $C_q =4 \pi \sigma'(0)/\epsilon=  4\pi e^2 n'(\mu)/\epsilon$ as the linear quantum capacitance. Effects of the electronic structure of the material (i.e. quantum effects) enter only through the quantity $C_q$ -- the remaining calculations will proceed using classical electrostatics. (In principle, our approach could equally apply in the high temperature or non-degenerate limit. Nonetheless, the systems considered here will typically be in the quantum limit.) With Equation (\ref{q_cap_lin}), the Gauss Law (\ref{G_law}) becomes a Robin boundary condition  
\begin{equation}
[\mathbf{n} \cdot \nabla \phi] =   - C_q \phi_s. 
\end{equation}
Note that we recover the classical boundary condition for an ideal metal, $\phi_s =0$, for $C_q = \infty$. 

We will exploit the superposition principle in the calculations below as the Robin boundary condition is linear. Taking into account higher order terms in the expansion of Equation (\ref{TF_nonlin}) would give non-linear terms in the boundary condition which would lead to a non-pairwise interionic interaction potential.

\section{Ions in a Slit Nanopore}
We now proceed to solve the electrostatic boundary value problem for a point charge in a slit nanopore (Figure \ref{schematic_nanotube}a). Noting the axisymmetry of the problem, the electric potential inside the pore satisfies the Poisson equation
\begin{equation}
\frac{1}{r} \frac{\partial}{\partial r} \left( r \frac{ \partial \phi_1}{\partial r} \right) + \frac{\partial^2 \phi_1 }{\partial z^2} = - \frac{4 \pi q}{\epsilon} \frac{\delta_r(r)}{2\pi r} \delta(z), \quad -\frac{L}{2}<z<\frac{L}{2}, 
\label{in_slit_laplace}
\end{equation} 
where $\delta_r(r)$ is the radial delta function and $\delta(z)$ is the 1D delta distribution. Outside the slit pore, the potential satisfies Laplace's equation 
\begin{equation}
\frac{1}{r} \frac{\partial}{\partial r} \left( r \frac{ \partial \phi_2}{\partial r} \right) + \frac{\partial^2 \phi_2 }{\partial z^2} = 0 , \quad  \frac{L}{2} < |z|, 
\label{out_slit_laplace}
\end{equation} 
boundary conditions for (\ref{in_slit_laplace}) -(\ref{out_slit_laplace}) are
\begin{align}
\phi_2\left(r,\pm \frac{L}{2} \right) = \phi_1 \left(r,\pm \frac{L}{2} \right) &\equiv \phi_s(r),  \label{cont_pot11} \\ 
\frac{\partial  \phi_2}{\partial z} \Bigg|_{z=\pm L/2} -  \frac{\partial  \phi_1}{\partial z} \Bigg|_{z=\pm L/2} &= C_q  \phi_s(r), 
\label{quant_cap1}
\end{align} 
with the condition that the potential decays away from point charge, 
\begin{equation}
\phi_{1,2}(r,z) \rightarrow 0, \; \mathrm{as} \; \sqrt{r^2 + z^2} \rightarrow \infty. 
\label{bdy_cond1}
\end{equation}
Introducing the Hankel transform 
\begin{equation}
\tilde{\phi}(k,z) = \int_{0}^{\infty} r J_0 (k r) \phi(z,r) \mathrm{d}r, 
\end{equation}
where $J_0(x)$ is the zeroth order Bessel function of the first kind, we obtain 
\begin{align}
\frac{\partial^2 \tilde{\phi}_1 }{\partial z^2} -k^2  \tilde{\phi}_1 &= - \frac{2 q}{\epsilon} \delta(z), \quad  -\frac{L}{2}<z<\frac{L}{2},  \\
\frac{\partial^2 \tilde{\phi}_2 }{\partial z^2} -k^2  \tilde{\phi}_2 &= 0 ,   \quad |z|>\frac{L}{2}.
\end{align} 
Noting the symmetry about $z=0$, the general solution is 
\begin{align}
\tilde{\phi}_1 &= - \frac{q}{\epsilon k} \sinh k|z| + B \cosh k z, \\ 
\tilde{\phi}_2 &=  C e^{- k |z|}. 
\end{align} 
The boundary conditions (\ref{quant_cap1}) - (\ref{bdy_cond1})  yield equations for the unknown constants $B$ and $C$, which are readily solved to give
\begin{align}
B &=  \frac{q}{\epsilon k} \frac{(C_q + k) \sinh \frac{k L}{2} + k \cosh \frac{k L}{2} }{(C_q + k) \cosh \frac{k L}{2} + k \sinh \frac{k L}{2} },\\
C &=   \frac{q}{\epsilon} e^{\frac{k L}{2}} \frac{1 }{ (C_q+k) \cosh \frac{kL}{2}  + k \sinh \frac{k L}{2} }. 
 \end{align} 
Therefore, performing an inverse Hankel transform, we obtain the central result of our calculations 
\begin{align}
\phi_{1}(r,z) &= -\frac{q}{\epsilon} \int_{0}^{\infty}\; J_0(k r) \left[ \sinh (k |z|)  \right. \nonumber \\ 
& \left. -  \frac{(C_q + k) \sinh \frac{k L}{2} + k \cosh \frac{k L}{2} }{(C_q + k) \cosh \frac{k L}{2} + k \sinh \frac{k L}{2} } \cosh k z\right]  \; \mathrm{d}k, \label{pot_in_slit}  \\
\phi_{2}(r,z) &=  \frac{q}{\epsilon} \int_{0}^{\infty}\;   \frac{ k e^{\frac{k L}{2}  } J_0(k r)  e^{- k |z|} }{ (C_q+k) \cosh \frac{kL}{2}  + k \sinh \frac{k L}{2} } \;  \mathrm{d}k. 
\label{pot_out_slit}
\end{align}

Equation (\ref{pot_in_slit}) shows that the electric potential, and thus the interionic interaction energy, decrease as the slit width decreases. Indeed, narrow pores with separations comparable to ion diameter are used in supercapacitors to minimise Coulombic repulsion between charges and thus maximise charge storage \cite{chmiola2006anomalous,largeot2008relation,lin2009solvent,kondrat2011superionic1,kondrat2014single}. In such close-fitting pores, ions are positioned on the axis of symmetry and the interaction energy between two ions is given by  
\begin{align}
\beta U(R) = \frac{l_B}{L}& \int_{0}^{\infty}  J_0(K R) \nonumber \\ 
&\times \frac{(\tilde{C}_q + K) \sinh \frac{K}{2} + K \cosh \frac{K}{2} }{(\tilde{C}_q + K) \cosh \frac{K}{2} + K \sinh \frac{K}{2} } \; \mathrm{d}K,  
\label{int_pot_slit}
\end{align}
where $\beta = 1/(k_B T)$, $l_B = q^2/(\epsilon k_B T)$ is the Bjerrum length, and we have introduced dimensionless quantities $R = r/L$, $K = k L$ and $\tilde{C}_q = C_q  L$. Noting the general asymptotic identity for integrals of product of smooth functions with Bessel functions (see \cite{willis1948lv})
\begin{equation}
\int_0^{\infty} J_0(m k) f(k) \mathrm{d}k = \frac{f(0)}{m} - \frac{1}{2} \frac{f''(0)}{m^3} + O(m^{-5}), 
\label{asym_hankel}
\end{equation}
the asymptotic behaviour of (\ref{int_pot_slit}) may immediately be determined. We find 
\begin{equation} 
U(R) \sim \frac{l_B}{L \tilde{C}_q^2} \frac{1}{R^3} + O(R^{-5}), \; \mathrm{as} \; R \rightarrow \infty, 
\end{equation} 
which is qualitatively different from the exponential decay predicted by ideal metal theory \footnote{In the limit $\tilde{C}_q \rightarrow \infty$, Equation (\ref{int_pot_slit}) tends to $U_{\mathrm{m}}(R) = \frac{l_B}{L} \int_{0}^{\infty}  J_0(K R) \tanh (K/2) \mathrm{d}K$. Equation (\ref{asym_hankel}) cannot be applied to extract the asymptotic behaviour of this integral as all even derivatives of $\tanh (K/2)$ vanishes at $K=0$, showing that the asymptotic decay is faster than all power laws. Using contour integration and noting the positions of the poles, the integral can rewritten as the sum $U_{\mathrm{m}}(R) = \frac{4 l_B}{L} \sum_{n=1}^{\infty} K_0\left( n \pi R/L \right)$. The first term of the sum is dominant as $R \rightarrow \infty$, and thus we obtain Equation (\ref{metal_asym}). } (recovered $\tilde{C}_q \rightarrow \infty$) 
\begin{equation} 
U_{\mathrm{m}}(R) \sim \frac{4 l_B}{L}\frac{1}{\sqrt{2 R}} e^{- \pi R} , \; \mathrm{as} \; R \rightarrow \infty. 
\label{metal_asym}
\end{equation} 
We note that Thomas-Fermi theory for bulk semimetals predicts a similar exponential decay albeit with a renormalised pore radius $R$ \cite{rochester2013interionic}.

\begin{figure}
\centering
\includegraphics[scale=0.25]{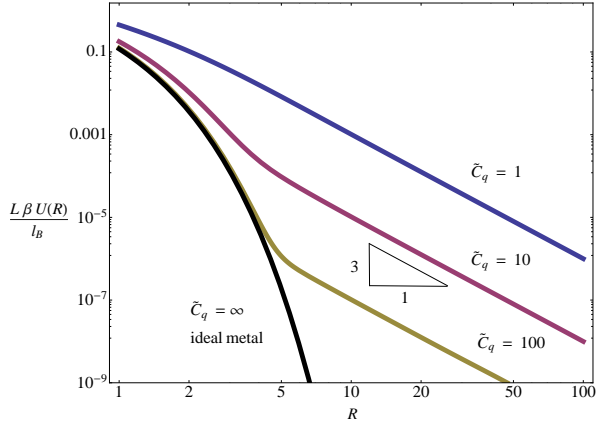}
\caption{The interionic interaction between two ions located on the symmetry plane ($Z=0$) as a function of the separation distance $R$ for different values of the quantum capacitance $\tilde{C}_q$. Note that $\tilde{C}_q=\infty$ corresponds to an ideally metallic pore. The curves are calculated by evaluating the integral (\ref{int_pot_slit}) numerically in \texttt{Mathematica}.}
\label{screening_slit}
\end{figure}

Figure \ref{screening_slit} confirms that the interionic potential decays $\sim 1/R^3$ as $R \rightarrow \infty$. Increasing the quantum capacitance $\tilde{C}_q$ screens the ion-ion interaction and decreases $U(R)$, but that does not alter the ultimate asymptotic decay behaviour of $U(R)$. For $\tilde{C}_q \gg 1$, the ion-ion interaction appears to decay exponentially for small $R$ before transitioning to algebraic decay when $R \gg 1$. Therefore, the confinement of the screening electrons to a 2D sheet qualitatively changes the interionic interactions. 

The measured quantum capacitance of pure graphene falls in the range 2-10 $\mathrm{\mu F \; cm^{-2}}$ \cite{xia2009measurement}. Assuming $L = 1 \mathrm{nm}$ and $\epsilon = 2$, in our dimensionless units $\tilde{C}_q \approx 1 - 5$. In this parameter regime, Figure \ref{screening_slit} shows that the deviation from ideal metallic behaviour is significant. 

\begin{figure}
\centering
\includegraphics[scale=0.25]{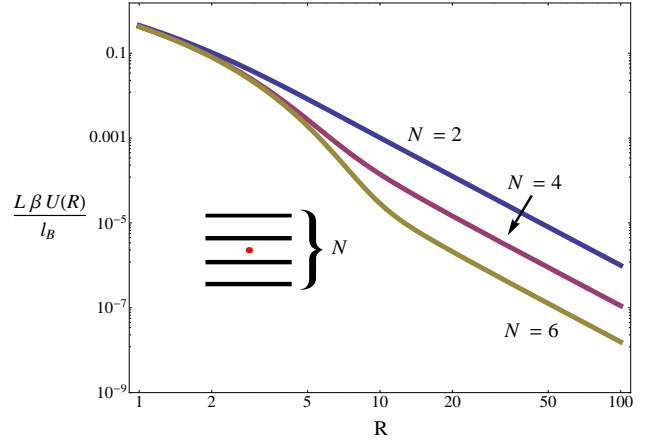}
\caption{The interionic interaction between two ions located on the symmetry plane ($Z=0$) of a graphene stack as a function of the separation distance $R$ for quantum capacitance $\tilde{C}_q=1$. The separations between the $N$ graphene sheets in the stack are assumed to be the same and equal to $L$. }
\label{stack}
\end{figure}

Our result, that graphene pores screen interionic interactions poorly, is robust even considering the effects of finite lateral extent of realistic carbon material and the presence of neighbouring pores. Realistic carbon materials are, of course, not infinite pores. Typical slit-like ordered domains of carbide-derived-carbon can range up to $\sim 20 \AA$ \cite{forse2015new}. Therefore, assuming a pore diameter $\sim 5 \AA$ (corresponding to a pore that is of the same size as typical ionic liquid ions), the algebraic decay regime is still well within the slit-like ordered domain. The effect of neighbouring pores can be evaluated by considering the interaction of ions in a graphene stack (see Figure \ref{stack}). In this stack geometry, Equation (\ref{in_slit_laplace}) still holds for $Z \in [-L/2,L/2]$, but Equation (\ref{out_slit_laplace}) needs to be solved in each outer region bounded by graphene sheets, and boundary conditions (\ref{cont_pot11})-(\ref{quant_cap1}) need to be imposed on each graphene surface. This furnishes a set of linear equations which can then be solved using \texttt{Mathematica}. We do not include the resulting expressions here as they are rather cumbersome. Figure \ref{stack} shows that the greater the number of graphene sheets in a stack, the weaker is the magnitude of the interionic interaction. However, the algebraic $\sim 1/r^3$ decay is robust and numerical data suggests that the prefactor $ \sim 1/ (\tilde{C}_q^2 N^2)$ for $\tilde{C}_q \ll 1$ and $\sim 1/\tilde{C}_q^{N}$ for $\tilde{C}_q \gg 1$ .

All in all, our calculations show that there is a crucial tradeoff between pores with thick semimetalic pore walls (\emph{e.g.} graphite pores) which deliver superior electronic properties but necessarily lower volumetric capacitance (as the pore walls occupy volume), and graphene-based pores which have inferior electronic properties but higher volumetric capacitance.

\section{Ions in a Cylindrical Nanotube}
Next we consider the electrostatic boundary-value problem for a point charge in a cylindrical nanotube (Figure \ref{schematic_nanotube}b). Noting the axisymmetry of the problem, the electric potential satisfies the Poisson equation
\begin{equation}
\frac{1}{r} \frac{\partial}{\partial r} \left( r \frac{ \partial \phi_1}{\partial r} \right) + \frac{\partial^2 \phi_1 }{\partial z^2} = - \frac{4 \pi q}{\epsilon} \frac{\delta_r(r)}{2\pi r} \delta(z), \quad 0<r<R. 
\end{equation} 
Outside the nanotube, the potential satisfies the Laplace equation 
\begin{equation}
\frac{1}{r} \frac{\partial}{\partial r} \left( r \frac{ \partial \phi_2}{\partial r} \right) + \frac{\partial^2 \phi_2 }{\partial z^2} = 0 ,  \quad r>R. 
\end{equation} 
The boundary conditions on the pore wall are
\begin{align}
\phi_2(a,z) = \phi_1(a,z) &\equiv \phi_s(z),  \label{cont_pot} \\ 
\frac{\partial  \phi_2}{\partial r} \Bigg|_{r=a} -  \frac{\partial  \phi_1}{\partial r} \Bigg|_{r=a} &=  C_q \phi_s(z), 
\label{quant_cap}
\end{align} 
complemented with the condition that the potential decays away from the point charge, 
\begin{equation}
\phi_{1,2}(r,z) \rightarrow 0, \; \mathrm{as} \; \sqrt{r^2 + z^2} \rightarrow \infty. 
\label{bdy_cond}
\end{equation}
Performing a Fourier cosine transform in the $z$-direction 
\begin{equation}
\hat{\phi}(k,r) = \int_{0}^{\infty} \cos (k z) \phi(z,r) \mathrm{d}z, 
\end{equation}
we obtain
\begin{align}
\frac{1}{r} \frac{\partial}{\partial r} \left( r \frac{ \partial \hat{\phi}_1}{\partial r} \right) -k^2  \hat{\phi}_1 &= - \frac{2 q}{\epsilon} \frac{\delta_r(r)}{r}, \quad 0<r<R, \label{ode_costrans} \\
\frac{1}{r} \frac{\partial}{\partial r} \left( r \frac{ \partial \hat{\phi}_2}{\partial r} \right) -k^2 \hat{\phi}_2 &= 0 ,  \quad r>R.
\label{ode_costrans1}
\end{align} 
Equations (\ref{ode_costrans})--(\ref{ode_costrans1}) have the general solution 
\begin{align}
\hat{\phi}_1 &= \frac{2q }{\epsilon \pi} K_0(k r) + B I_0 (k r), \label{gsol} \\ 
\hat{\phi}_2 &= C K_0(k r) + D I_0 (k r),  \label{gsol1} 
\end{align} 
where $I_0(x)$ and $K_0(x)$ are zeroth order modified Bessel functions of the first and second kind, respectively. As the potential must decay away from the point charge (\emph{c.f.} Equation (\ref{bdy_cond})), $D = 0$. The other constants, $B$ and $C$, can be determined by the continuity of the potential across $r=a$ (\ref{cont_pot}), and the quantum capacitance condition (\ref{quant_cap}). These conditions yield
\begin{align}
B & =- \frac{2 q}{\epsilon \pi} \frac{ C_q a  K_0(k a)^2}{1 + C_q a I_0(k a) K_0(k a) }, \\
C & = \frac{2 q}{\epsilon \pi} \frac{1}{1 + C_q a I_0(k a) K_0(k a)}. 
\end{align} 
After performing an inverse cosine transform on Equations (\ref{gsol})-(\ref{gsol1}), the electric potential inside and outside the slit is given by 
\begin{align}
\phi_{1}(r,z) &= \frac{q}{\epsilon \sqrt{r^2 + z^2}} \nonumber \\ 
&-  \frac{2q}{\epsilon \pi} \int_0^{\infty} \; \frac{ C_q a  K_0(k a)^2}{1 + C_q a I_0(k a) K_0(k a) } \; I_0(kr) \cos k z \; \mathrm{d}k, \label{pot_in_cyl}  \\
\phi_{2}(r,z) &= \frac{2q}{\epsilon \pi} \int_{0}^{\infty} \frac{1}{1 + C_q a I_0(k a) K_0(k a)} K_0(k r) \cos kz \; \mathrm{d}k.
\label{pot_out_cyl}
\end{align}

\begin{figure}
\centering
\includegraphics[scale=0.25]{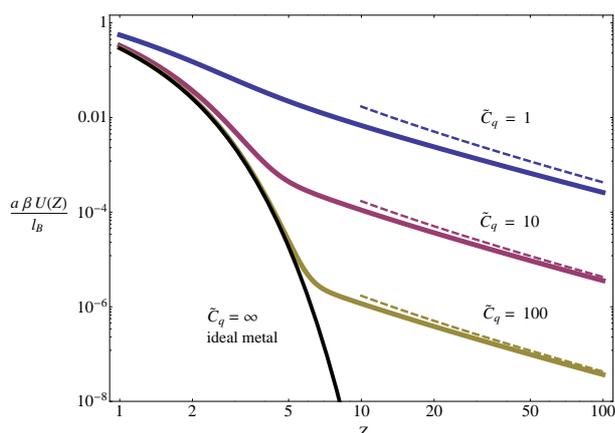}
\caption{The interionic interaction between two ions located at the central symmetric axis as a function of separation distance $Z$ for different quantum capacitance $C_q$. The dotted lines show the asymptotic result, Equation (\ref{asym_pore}).}
\label{screening_cyl}
\end{figure}

In particular, the interionic interaction between two ions located on the symmetry axis ($r=0$) of the pore is given by 
\begin{equation}
\beta U(Z) = \frac{l_B}{a Z} - \frac{2}{\pi} \frac{l_B}{a}  \int_0^{\infty} \; \frac{\tilde{C}_q K_0(K)^2}{1 + \tilde{C}_q I_0(K) K_0(K)}  \cos K Z \; \mathrm{d}K,
\label{pot_cyl}
\end{equation} 
where we have introduced dimensionless quantities $K = ka$, $Z = z/a$ and $\tilde{C}_q = C_q a$. The large $Z$ behaviour of (\ref{pot_cyl}) cannot be determined by an asymptotic expansion similar to (\ref{asym_hankel}) as the integrand in (\ref{pot_cyl}) is non-analytic around $K=0$. Instead, we note that the electric potential of a point charge in an ideally metallic pore ($\tilde{C}_q \rightarrow \infty$) is given by \cite{rochester2013interionic}
\begin{align} 
\beta U_{\mathrm{m}}(Z) &=  \frac{l_B}{a Z} - \frac{2}{\pi} \int_{0}^{\infty} \frac{K_0(K)}{I_0(K)} \cos KZ \; \mathrm{d}K \nonumber \\ 
& =  \frac{2 l_B}{a} \sum_{m=0}^{\infty} \frac{e^{-k_m Z}}{k_m J_1(k_m)^2} \sim e^{-k_0 Z} \;\; Z \gg 1,
\end{align} 
where $k_m$ is the $m^{th}$ root of $J_0(x)=0$. Thus we can isolate the exponentially decaying contribution in the potential from the long-ranged contributions  
\begin{align}
\beta U(Z) &=  \frac{2 l_B}{a} \sum_{m=0}^{\infty} \frac{e^{-k_m Z}}{k_m J_1(k_m)^2}  \nonumber \\ 
&+  \frac{2}{\pi} \frac{l_B}{a}  \int_0^{\infty} \; \frac{ K_0(K)}{I_0(K) +  \tilde{C}_q I^2_0(K) K_0(K)}  \cos K Z \; \mathrm{d}K.
\label{pot_cyl_1}
\end{align} 
The large $Z$ behaviour of (\ref{pot_cyl_1}) is determined by the small-$K$ expansion of the non-oscillatory part of the integrand, 
\begin{equation}
\frac{ K_0(K)}{I_0(K) +  \tilde{C}_q I^2_0(K) K_0(K)} \sim \frac{1}{\tilde{C}_q} + \frac{1}{\tilde{C}_q^2} \frac{1}{\log(1/K)} + \cdots 
\end{equation}
Noting that as $Z \rightarrow \infty $, 
\begin{align}
 &\int_0^{\infty} \; \frac{ K_0(K)}{I_0(K) +  \tilde{C}_q I^2_0(K) K_0(K)}  \cos K Z \; \mathrm{d}K \nonumber \\ 
 &\sim \int_0^{2\pi/Z} \;  \left( \frac{1}{\tilde{C}_q}  + \frac{1}{\tilde{C}_q^2} \frac{1}{\log(1/K)} \right) \cos K Z  \mathrm{d}K \\ \nonumber 
 & =  \frac{1}{\tilde{C}_q^2 Z } \int_0^{2\pi} \frac{\cos p}{\log (Z/p)} \mathrm{d}p \\ \nonumber 
 & \sim \frac{1}{\tilde{C}_q^2 Z \log Z }  \int_0^{2\pi} \cos p \left( 1+ \frac{\log p}{\log Z} \right) \mathrm{d}p  \\ \nonumber 
 & = \frac{\mathrm{Si}(2 \pi)}{\tilde{C}_q^2 Z \log^2 Z } 
\end{align} 
where $\mathrm{Si}(x)$ is the sine integral \cite{abramowitz1965handbook}. Note that the integral range to $\infty$ is replaced by $2 \pi/Z$ as the asymptotic behaviour is dominated by the first period of $\cos K Z$, the oscillatory part of the integrand. Therefore, the potential decays algebraically for large $Z$ and 
\begin{equation}
\beta U(Z) \sim 1.42\times \frac{2 l_B}{\pi a \tilde{C}_q^2} \frac{1}{Z \log^2 Z}. 
\label{asym_pore}
\end{equation} 
Figure \ref{screening_cyl} confirms that the interaction potential of ions in a nanotube $\propto 1/(Z \log^2 Z)$ for large ion separations, qualitatively different from the exponential decay for ideal metallic pores. Similar to the behaviour for slit pores, for $\tilde{C}_q \gg 1$ the ion-ion interaction appears to decay exponentially for small $R$ before transiting to algebraic decay when $R \gg 1$. This exponential decay for small ion separations agrees with quantum density functional theory calculations \cite{mohammadzadeh2014nanotubes,goduljan2014screening} for short carbon nanotubes. However, Equation (\ref{asym_pore}) suggests that results from simulations of short carbon nanotubes do not reflect the asymptotic decay behaviour. Indeed, the quantum capacitance of a $(6,6)$ single-walled carbon nanotube, estimated using density functional theory, is $\sim 9 \; \mathrm{\mu F \; cm^{-2}} $ \cite{pak2013relative}. In our dimensionless units, this corresponds to $\tilde{C}_q = 2$ (taking $a=0.4 \mathrm{nm}$ and as in the case for slit pores assume $\epsilon =2$). As the typical length of nanotubes is $O(100 \; \mathrm{nm})$ \cite{streit2012measuring}, the $\propto 1/(Z \log^2 Z)$ decay of interionic interaction is significant for realistic systems. We neglect the effects of neighbouring nanotubes here --- extrapolating from our result for a graphene stack (Figure \ref{stack}), the presence of a nanotube forest will likely weaken the magnitude of the interionic interaction, but crucially not its dependence on the interionic separation. 

The low quantum capacitance of carbon nanotubes suggests that conducting metallic nanopores with thick pore walls \cite{kim2008fabrication,buttner2009formation,bian2011ultrasmall,esterle2012evidence}, with an exponential decay of charge-charge interactions, may be far superior as materials for supercapacitors than nanopores made with carbon nanotubes. 

Previous theoretical models of charge storage in nanopores \cite{kondrat2014single,schmickler2015simple,rochester2015statistical} rely on the simplifying assumption that considering nearest-neighbour interactions are sufficient to account for electrostatic interactions between ions. While this assumption holds for exponentially decaying interactions (see e.g. qualitative comparison with Monte Carlo simulations in \cite{kondrat2014single}), with a long-ranged ion-ion interaction this assumption is suspect. In particular, results for the absence of phase transition in one dimension only holds for short-ranged interactions \cite{cuesta2004general}, thus charge storage in carbon nanotubes may proceed via phase transition akin to slit nanopores \cite{kondrat2011superionic1}. 

\section{Conclusion}
By including a finite quantum capacitance, we have shown that the screening of ion-ion interactions confined in semiconducting nanopores is far inferior to metallic nanopores. For a graphene slit pore, the interaction decays asymptotically $\propto 1/r^3$, whereas in a carbon nanotube, the asymptotic interaction potential is $\propto 1/(r \log^2 r)$, with $r$ the ion-ion separation. Hitherto works on capacitance optimisation have studied the effect of ion size and pore size, but our calculations highlight that the electronic structure of the pore walls themselves should be taken into account. To optimize capacitance, one should move away from pores based on semiconducting 2D materials and instead use metallic nanopores or bulk semimetallic materials.

\acknowledgments
This work was supported by an EPSRC Research Studentship and Fulbright Scholarship (AAL) and by the European Research Council (Starting Grant GADGET No. 637334 to DV).

\bibliographystyle{eplbib}
\bibliography{screening} 

\begin{thebibliography}{10}
\expandafter\ifx\csname url\endcsname\relax\def\url#1{\texttt{#1}}\fi

\bibitem{brandt2013ionic}
\Name{Brandt A., Pohlmann S., Varzi A., Balducci A. \and Passerini S.}
  \REVIEW{MRS bulletin}{38}{2013}{554}.

\bibitem{fujimoto2013electric}
\Name{Fujimoto T. \and Awaga K.} \REVIEW{Physical Chemistry Chemical
  Physics}{15}{2013}{8983}.

\bibitem{chmiola2006anomalous}
\Name{Chmiola J., Yushin G., Gogotsi Y., Portet C., Simon P. \and Taberna P.}
  \REVIEW{Science}{313}{2006}{1760}.

\bibitem{largeot2008relation}
\Name{Largeot C., Portet C., Chmiola J., Taberna P., Gogotsi Y. \and Simon P.}
  \REVIEW{J. Am. Chem. Soc.}{130}{2008}{2730}.

\bibitem{lin2009solvent}
\Name{Lin R., Huang P., Segalini J., Largeot C., Taberna P.-L., Chmiola J.,
  Gogotsi Y. \and Simon P.} \REVIEW{Electrochimica Acta}{54}{2009}{7025}.

\bibitem{kondrat2011superionic1}
\Name{Kondrat S. \and Kornyshev A.} \REVIEW{Journal of Physics: Condensed
  Matter}{23}{2011}{022201}.

\bibitem{skinner2011theory}
\Name{Skinner B., Chen T., Loth M. \and Shklovskii B.} \REVIEW{Physical Review
  E}{83}{2011}{056102}.

\bibitem{kondrat2011superionic}
\Name{Kondrat S., Georgi N., Fedorov M.~V. \and Kornyshev A.~A.}
  \REVIEW{Physical Chemistry Chemical Physics}{13}{2011}{11359}.

\bibitem{kondrat2014single}
\Name{Lee A.~A., Kondrat S., Kornyshev A.~A. \etal} \REVIEW{Physical Review
  Letters}{113}{2014}{048701}.

\bibitem{jackson1962classical}
\Name{Jackson J.~D. \and Jackson J.~D.} \Book{Classical electrodynamics} Vol.~3
  (Wiley New York etc.) 1962.

\bibitem{skinner2011model}
\Name{Skinner B., Fogler M. \and Shklovskii B.} \REVIEW{Physical Review
  B}{84}{2011}{235133}.

\bibitem{rochester2013interionic}
\Name{Rochester C.~C., Lee A.~A., Pruessner G. \and Kornyshev A.~A.}
  \REVIEW{ChemPhysChem}{14}{2013}{4121}.

\bibitem{mohammadzadeh2014nanotubes}
\Name{Mohammadzadeh L., Goduljan A., Juarez F., Quaino P., Santos E. \and
  Schmickler W.} \REVIEW{Electrochimica Acta}{162}{2015}{11}.

\bibitem{ghaznavi2010nonlinear}
\Name{Ghaznavi M., Mi{\v{s}}kovi{\'c} Z. \and Goodman F.} \REVIEW{Physical
  Review B}{81}{2010}{085416}.

\bibitem{willis1948lv}
\Name{Willis H.} \REVIEW{The London, Edinburgh, and Dublin Philosophical
  Magazine and Journal of Science}{39}{1948}{455}.

\bibitem{xia2009measurement}
\Name{Xia J., Chen F., Li J. \and Tao N.} \REVIEW{Nature
  nanotechnology}{4}{2009}{505}.

\bibitem{forse2015new}
\Name{Forse A.~C., Merlet C.~l., Allan P.~K., Humphreys E.~K., Griffin J.~M.,
  Aslan M., Zeiger M., Presser V., Gogotsi Y. \and Grey C.~P.}
  \REVIEW{Chemistry of Materials}{27}{2015}{6848}.

\bibitem{abramowitz1965handbook}
\Name{Abramowitz M. \and Stegun I.~A.} \Book{Handbook of mathematical functions
  with formulas, graphs, and mathematical tables} (Dover Publications) 1965.

\bibitem{goduljan2014screening}
\Name{Goduljan A., Juarez F., Mohammadzadeh L., Quaino P., Santos E. \and
  Schmickler W.} \REVIEW{Electrochemistry Communications}{45}{2014}{48}.

\bibitem{pak2013relative}
\Name{Pak A.~J., Paek E. \and Hwang G.~S.} \REVIEW{Physical Chemistry Chemical
  Physics}{15}{2013}{19741}.

\bibitem{streit2012measuring}
\Name{Streit J.~K., Bachilo S.~M., Naumov A.~V., Khripin C., Zheng M. \and
  Weisman R.~B.} \REVIEW{ACS nano}{6}{2012}{8424}.

\bibitem{kim2008fabrication}
\Name{Kim M., Jeong G.~H., Lee K.~Y., Kwon K. \and Han S.~W.} \REVIEW{Journal
  of Materials Chemistry}{18}{2008}{2208}.

\bibitem{buttner2009formation}
\Name{Buttner C.~C., Langner A., Geuss M., Muller F., Werner P. \and Gosele U.}
  \REVIEW{ACS nano}{3}{2009}{3122}.

\bibitem{bian2011ultrasmall}
\Name{Bian F., Tian Y., Wang R., Yang H., Xu H., Meng S. \and Zhao J.}
  \REVIEW{Nano letters}{11}{2011}{3251}.

\bibitem{esterle2012evidence}
\Name{Esterle T.~F., Sun D., Roberts M.~R., Bartlett P.~N. \and Owen J.~R.}
  \REVIEW{Physical Chemistry Chemical Physics}{14}{2012}{3872}.

\bibitem{schmickler2015simple}
\Name{Schmickler W.} \REVIEW{Electrochimica Acta}{173}{2015}{91}.

\bibitem{rochester2015statistical}
\Name{Rochester C.~C., Pruessner G. \and Kornyshev A.~A.}
  \REVIEW{Electrochimica Acta}{174}{2015}{978}.

\bibitem{cuesta2004general}
\Name{Cuesta J.~A. \and S{\'a}nchez A.} \REVIEW{Journal of statistical
  physics}{115}{2004}{869}.

\end{thebibliography}

\end{document}